# Phase Transition in the Three-Dimensional $\pm J$ Ising Spin Glass

N. Kawashima*

*Department of Physics, Toho University, Miyama 2-2-1, Funabashi, Chiba 274, Japan*

A. P. Young

*Department of Physics, University of California, Santa Cruz, CA 95064*

(October 2, 1995)

We have studied the three-dimensional Ising spin glass with a $\pm J$ distribution by Monte Carlo simulations. Using larger sizes and much better statistics than in earlier work, a finite size scaling analysis shows quite strong evidence for a finite transition temperature, $T_c$, with ordering below $T_c$. Our estimate of the transition temperature is rather lower than in earlier work, and the value of the correlation length exponent, $\nu$, is somewhat higher. Because there may be (unknown) corrections to finite size scaling, we do not completely rule out the possibility that $T_c = 0$ or that $T_c$ is finite but with no order below $T_c$. However, from our data, these possibilities seem less likely.

The question of whether there is a finite transition temperature, $T_c$, in an Ising spin glass in three dimensions has aroused a lot of interest for the last two decades[1], and the consensus of opinion has changed several times. About one decade ago, several pieces of work[2-5] seemed to show that there is a finite $T_c$, and this conclusion has generally been restated since then[6]. However, on closer inspection, the case is not completely closed. For example, the work of one of us[2], henceforth referred to as BY, is unable to rule out the possibility that $T_c = 0$ and the correlation length, $\xi$, diverges *exponentially* as $T \to 0$, as happens in the two-dimensional Heisenberg ferromagnet. The data is also consistent with a line of critical points terminating at $T_c \simeq 1.2$, as occurs in the Kosterlitz-Thouless-Berezinskii theory of the two-dimensional XY ferromagnet. In this scenario there would be no long range spin glass order below $T_c$. Furthermore, recent results of Marinari et al.[7] were found to be consistent *both* with a finite $T_c$ and with a zero temperature transition where the correlation length diverges exponentially, $\xi \sim \exp(A/T^4)$. We therefore feel there are three possible scenarios, consistent with existing work:

(i) $T_c$ is finite and there is spin glass order at lower temperatures,

(ii) $T_c$ is finite but there is a line of critical points (i.e. no spin glass order) at lower temperatures,

(iii) $T_c = 0$ and the correlation length diverges exponentially as $T \to 0$.

During the last decade available computer power has increased enormously so, given these uncertainties, it is useful to look at the problem again. The calculations presented here are similar to those of BY, but we are able to study larger system sizes in the temperature range of interest and obtain *much* better statistics by averaging over many more samples. As a result, unlike BY, we are able to see clear evidence for *ordering below a finite $T_c$*.

The Hamiltonian is

$$\mathcal{H} = -\sum_{\langle i,j \rangle} J_{ij} S_i S_j ,  \quad (1)$$

TABLE I. For each size, $L$, we show the largest value of $t_0$, (where, as explained in the text, the simulation ran for $3t_0$ sweeps) and the minimum number of samples, $N_s$.

| L | largest $t_0$ | minimum $N_s$ |
|---|---|---|
| 6 | $4 \times 10^5$ | 8192 |
| 8 | $1 \times 10^6$ | 8192 |
| 12 | $8 \times 10^6$ | 6880 |
| 16 | $15 \times 10^6$ | 3392 |
| 24 | $5 \times 10^6$ | 2080 |

where the spins $S_i$ take values $\pm 1$, and the nearest neighbor interactions, $J_{ij}$ take values $\pm 1$ with equal probability. The simple cubic lattice contains $N = L^3$ spins and has periodic boundary conditions. In some previous work, $\{J_{ij}\}$ was generated so that the the number of ferromagnetic couplings is exactly the same as that of antiferromagnetic couplings. We do not impose such a condition in the present work.

The Monte Carlo simulation uses a multispin coding technique[8] in which each spin and bond is represented by a single bit of a computer word. On a 32 bit machine we then flip in parallel 32 spins (on the same lattice site but in different samples with different realizations of the disorder). For this method to be efficient the *same* random number is used for each bit[9]. We use a shift register random number generator[10,11], commonly known as R250. The code runs at 27 million spin updates per second on one node (IBM 390 RISC workstation) of the SP2 computer at the Maui High Performance Computing Center. Since we need many more than 32 samples, we ran the same code independently on many nodes at the same time. Each node produces its own output file from which the final averaging is easily done using a unix shell script. Monte Carlo simulations of random systems thus provide an example where parallel computing can be done in a trivial (and almost perfectly efficient!) way. The total CPU time used for the data presented here is about 9 node–years.

To get good statistics we average over a large number of samples, $N_s$, where for each size, $N_s$ is at least the value in the third column in table I. After $t_0$ sweeps for equilibration, an additional $2t_0$ sweeps are carried out for measurements. For each size, the largest value of $t_0$ used is also shown in table I (this is for the lowest temperature: at higher temperatures many fewer sweeps are generally needed).

As usual[2], for each realization of the bonds, two copies of the system are studied with different initial values of the spins and different random numbers for generating the spin flips. Of particular importance is the overlap between the two copies,

$$q = \frac{1}{N} \sum_{i=1}^{N} S_i^{(1)} S_i^{(2)} , \quad (2)$$

where the labels "1" and "2" denote the copies. From measurements of $q$ we compute the Binder ratio[12,2]

$$g = \frac{1}{2}\left[3 - \frac{\langle q^4 \rangle}{\langle q^2 \rangle^2}\right] , \quad (3)$$

where the average $\langle \ldots \rangle$ denotes *both* a thermal average for a given set of bonds and an average over the disorder[13]. At high temperature, $g \to 0$, whereas $g \to 1$ in the spin glass phase, at least if there is a unique thermodynamic state.

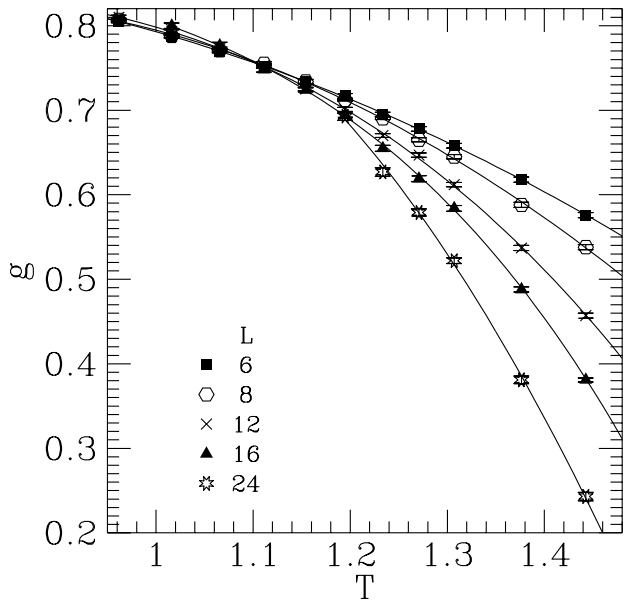

FIG. 1. Results for the Binder ratio $g$, defined in Eq. (3), for different sizes and temperatures. The lines are smooth curves through the data and are only intended as guides to the eye.

Because $g$ is dimensionless it has the finite size scaling form[2]

$$g = \tilde{g}(L^{1/\nu}(T - T_c)) \quad (4)$$

and so is *independent of $L$* at $T_c$. The behavior of $g$ is different for each of the three scenarios discussed above:

(i) the curves for $g$ will intersect at $T_c$ and splay out again at lower $T$ (with the larger sizes having the larger values, the opposite of the situation above $T_c$),

(ii) the curves for $g$ will come together at $T_c$ and then stick together at lower $T$,

(iii) the curves will merge once $\xi \gg L$, but data for larger sizes will merge to this common curve at successively lower temperatures.

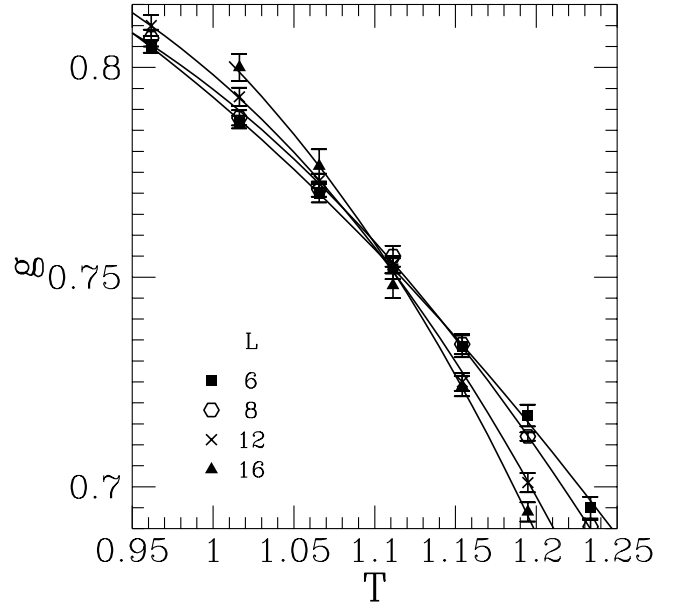

FIG. 2. An enlarged view of the data in Fig. 1 in the crucial region where the curves come together.

In addition to $g$, we also computed the spin glass susceptibility,

$$\chi_{SG} = N \langle q^2 \rangle , \quad (5)$$

and $P(q)$, the distribution of $q$. These have the finite size scaling forms,[2]

$$\chi_{SG}(L^{1/\nu}(T - T_c)) , \quad (6)$$

and

$$P(q) = L^{\beta/\nu} \tilde{P}(L^{\beta/\nu} q, L^{1/\nu}(T - T_c)) , \quad (7)$$

where $\beta$ is the order parameter exponent and is related to $\eta$, which gives the power law decay of the correlations at the critical point, by

$$\frac{\beta}{\nu} = \frac{1}{2}(d - 2 + \eta) . \quad (8)$$

It is very important to ensure that enough Monte Carlo sweeps have been carried out to equilibrate the sample. Following BY we compare the results for $g$ obtained, as described above, from the overlap between two replicas with the results obtained from one replica at two different times (see BY for details). BY found that these two estimates approach the equilibrium value from opposite directions as the length of the simulation increased. Once the two values agree, they do not change further if more sweeps are carried out. We have also tested this by doing the run for $L = 8, T = 0.9618$ for an order of magnitude longer time than needed for the two estimates to agree. Again we find that there is no subsequent change within our (much smaller) errors.

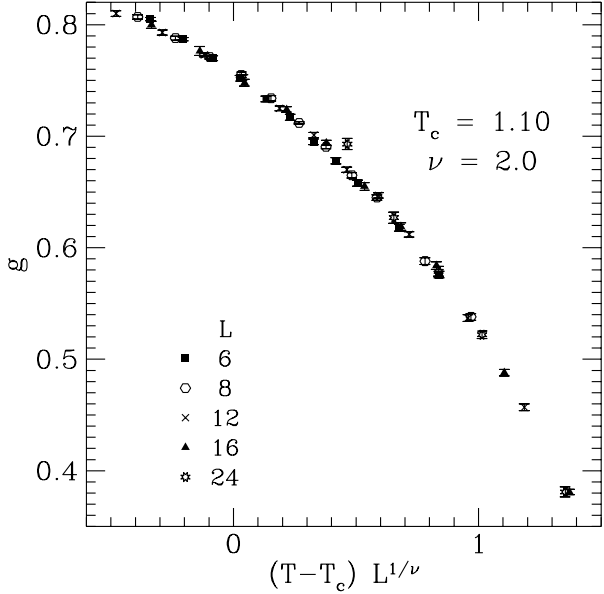

FIG. 3. A scaling plot for $g$ according to the form in Eq. (4).

Our data for $g$ is shown in Fig. 1 and an enlarged view of the region where the curves for different sizes intersect is shown in Fig. 2. From Fig. 2 one sees clear evidence for splaying out of the data below a temperature of about 1.10. Estimating $T_c$ to be approximately 1.10 from the intersection point we can scale most of the data according to Eq. (4) with $\nu = 2.0$, see Fig. 3. The only point which does not lie on a common curve is the result for $L = 24, T = 1.1948$, which is significantly higher. One can see from Fig. 1 that this point has almost the same value of $g$ as the data for $L = 16$ at the same temperature. This data point being rather higher than expected *may* reflect corrections to finite size scaling, and indicate that the true critical temperature is higher than the straightforward estimate based on data for $g$ with $L \leq 16$.

Once $T_c$ has been estimated, one can obtain $\beta/\nu$, or equivalently $\eta$, from the expected scaling form of $P(q)$ at criticality given by Eq. (7) with $T = T_c$. The corresponding plot is shown in Fig. 4 for $T = 1.1113$ (well within the bounds of our estimate of $T_c$), and has $\beta/\nu = 0.3$ which corresponds to $\eta = -0.4$ from Eq. (8) with $d = 3$.

We have also performed finite size scaling plots for $\chi_{SG}$ according to Eq. (6). This data does not locate $T_c$ precisely, so we have used the same $T_c$ as obtained from the scaling plot for $g$ in Fig. 3, i.e. $T_c = 1.10$. Furthermore the value of $\eta$ is constrained by requiring that the data scales at $T_c$ and from Fig. 4 this gives $\eta = -0.4$ The only remaining parameter is $\nu$ and the best fit, shown in Fig. 5, is for $\nu = 1.6$.

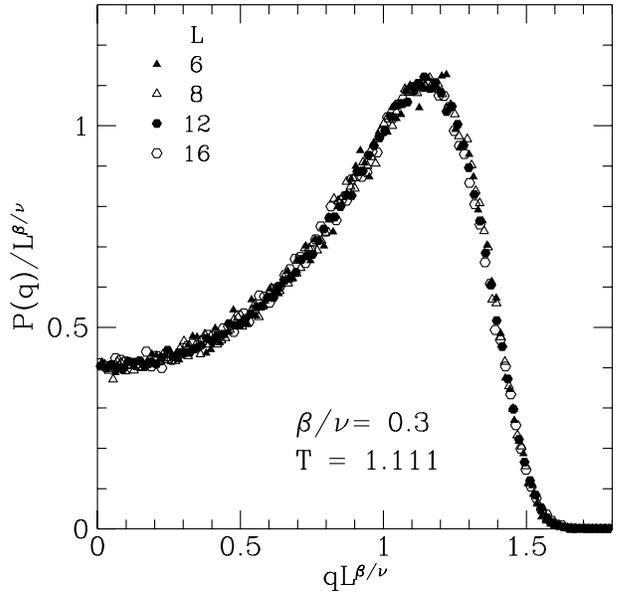

FIG. 4. A scaling plot for $P(q)$ at $T = 1.1113$ (which is close to the the critical point) according to the form in Eq. (7). According to Eq. (8), the value $\beta/\nu = 0.3$ corresponds to $\eta = -0.4$.

The values for $\nu$ obtained from $g$ and $\chi_{SG}$ are somewhat different. If we try to use $\nu = 2.0$ in the data for $\chi_{SG}$ or $\nu = 1.6$ in the data for $g$, the fit is visibly worse. Presumably this difference indicates that corrections to finite size scaling are not negligible for the range of sizes that we can study. Taking into account all the data we estimate

$$T_c = 1.11 \pm 0.04$$
$$\nu = 1.7 \pm 0.3$$
$$\eta = -0.35 \pm 0.05 \ . \quad (9)$$

As discussed above, the $L = 24$ data indicates that $T_c$ may be higher than that estimated from the intersections of $g$ for $L \leq 16$. This is reflected in the estimated error for $T_c$ in Eq. (9). The estimated errors in $\nu$ and $\eta$ then come largely from the uncertainty in $T_c$. Our value of $T_c$ is rather lower than earlier estimates which were close to 1.2, and the value of $\nu$ is higher, previous estimates generally being in the vicinity of 1.3. Our value of $\eta$ is not very different from earlier estimates.

To conclude, we have found evidence for a finite transition temperature with spin glass order below $T_c$, scenario (i) above. However, it is difficult to estimate the size of systematic errors, such as possible correlations in the random numbers (though we believe that these are very small[11]), and corrections to finite size scaling. Because of this, and because the crossing of the data for $g$ that we observe in Fig 2 is rather small, we cannot rule out, for sure, the other two possibilities, i.e. scenario (ii) in which $T_c$ is finite but there is no spin glass order at lower temperature, or scenario (iii) in which $T_c = 0$. However, from our data, these possibilities now seem less likely.

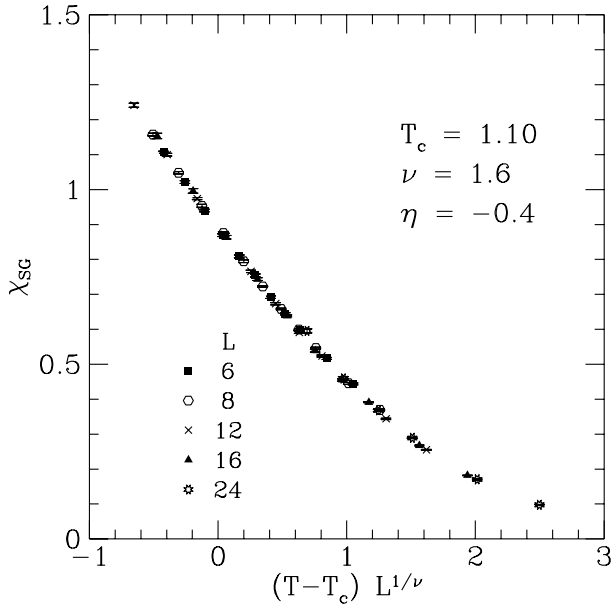

FIG. 5. A scaling plot for $\chi_{SG}$ according to the form in Eq. (6).

Since the present study required a substantial computer effort, an investigation of larger sizes, which is necessary to confirm scenario (i) beyond reasonable doubt, may need a better algorithm than single spin flip Monte Carlo. There are already some promising results from the "replica exchange" method[14] (where, in addition to local moves, global moves are made which cause the temperature of the system to cycle up and down).

We would like to thank K. Nemoto, K. Hukushima, H. Takayama and P. Coddington for interesting discussions. One of us (APY) would also like to thank D. Stauffer for helpful comments. This work has been supported by the National Science Foundation under grant No. DMR–9411964. We would also like to thank the Maui High Performance Computing Center for a generous allocation of computer time.